\documentclass[iop]{emulateapj}

\usepackage{ifsym}
\usepackage{amssymb}
\usepackage{amsmath}
\usepackage{graphicx}
\usepackage[hypertex, colorlinks=true, linkcolor=blue, citecolor=blue, urlcolor=blue]{hyperref}
\usepackage{color}

\shorttitle{Resonant Amplification of Turbulence by Blast Waves}
\shortauthors{Zankovich and Kovalenko}

\begin{document}

\title{Resonant Amplification of Turbulence by the Blast Waves}

\author{A. M. Zankovich and I. G. Kovalenko}
\affil{Physicotechnical Institute, Volgograd State University, Volgograd 400062, Russia}
\email{ilya.g.kovalenko@gmail.com}

\begin{abstract}
We discuss an idea whether spherical blast waves can amplify by a non-local resonant hydrodynamic mechanism inhomogeneities formed by turbulence or phase segregation in the interstellar medium. We consider the problem of a blast-wave-turbulence interaction in the Linear Interaction Approximation.
Mathematically, this is an eigenvalue problem for finding the structure and amplitude of eigenfunctions describing the response of the shock-wave flow to forced oscillations by external perturbations in the ambient interstellar medium. Linear analysis shows that the blast wave can amplify density and vorticity perturbations for a wide range of length scales with amplification coefficients of up to 20, with amplification the greater, the larger the length. There also exist resonant harmonics for which the gain becomes formally
infinite in the linear approximation. Their orbital wavenumbers are within the range of macro- ($l \sim 1$), meso- ($l \sim 20$) and microscopic ($l > 200$) scales. Since the resonance width is narrow: typically, $\Delta l <1$, resonance should select and amplify discrete isolated harmonics. We speculate as to a possible explanation of an observed regular filamentary structure of regular-shaped round supernova remnants such as  SNR 1572, 1006 or 0509-67.5. Resonant mesoscales found ($l \approx 18 $) are surprisingly close to the observed scales ($l \approx 15$) of ripples in the shell's surface of SNR 0509-67.5.
\end{abstract}


\section{INTRODUCTION}

The shapes of supernova remnants (SNRs) are generally far from perfect spherical geometry. Various physical factors cause asphericity, among these are anisotropy of supernova ejecta, hydrodynamic instabilities or inhomogeneities in an ambient interstellar medium. Even the most rotund remnants such as SNR 1572 and 1006 \citep{Raymond07,Winkler13} or SNR 0509-67.5 \citep{Warren04} are endowed with small scale structures such as ripple, filaments or knots.

The latter two have corrugations with a radius of curvature less than or equal to one tenth of the SNR size \citep{Raymond07}. \citet{Raymond03} and \citet{Raymond07} argue for the turbulent origin of the rippling in SNR 1006 as well as in the other filamentary SNR Cygnus Loop while \citet{Patnaude02} speak for multiphaseness.

Even accepting the hypothesis of turbulent or multiphase origin we have to admit at the same time an amazing regularity of pattern imprinting fine rippling that can be seen on the recently published composite optical-X-ray image of SNR 0509-67.5 from {\sl Hubble} and {\sl Chandra} released by \citet{NASA12}. One can recognize $\sim 15$ knots observed in longitude: this regular speckled structure envelops the whole remnant's surface. This regular pattern suggests that some lengthscales of perturbations are more preferred than others and this is hard to understand in a turbulence scenario that is scaleless by nature.

Quite the reverse, it would appear reasonable that there exists a physical mechanism amplifying few intermediate scales. A plausible explanation would be that we observe the development of some instability on the shock surface of the remnant. If the idea of instability is correct, the fact that the excited scales are not vanishingly small but comparable to the remnant's size argues for the non-local nature of instability. This means that different fragments of the filamentary structure are physically coupled, which provides for ordering and regularity of the filamentary structure irrespective of initial and, probably, environmental conditions. In the case in point the globality of instability can be caused by the remnant's size finiteness.

There are a multitude of different types of shock wave instabilities among which the global instabilities constitute a modest set.
Regarding the SNR shocks instabilities the case that should be mentioned above all else is the Ryu-Vishniac instability \citep{RV87,RV91}. This instability is based on the hydrodynamic mechanism of self-excitation of oscillations within a spherical shock wave. However, its scope is limited by gases in almost isothermal state, the increments are sufficiently small, and saturation occurs at weakly non-linear level (the density enhancements have a factor of $\lesssim 2$) \citep{MacLow93}.

The spherical shock in the Ryu-Vishniac representation can be considered a coherent oscillator generating self-sustained oscillations. The spherical shock again is a spherical resonant cavity that can be excited by external disturbances.
Indeed, the environment for the interstellar shock is not in the least a perfect uniform background but usually a strongly inhomogeneous, turbulized or phase-separated medium.

In the present paper we analyze the efficiency of possible resonant amplification of disturbances by the spherical shock wave. The basic idea is that a supernova remnant can filter out some resonant length scales within chaotic and globally scaleless noise in the ambient turbulent or phase-separated interstellar matter.

The possibility of turbulence amplification by the shock waves has been a debatable issue for a long time. Such a possibility has been proved theoretically \citep{Wouchuk,Donzis}, numerically \citep{Lee97,Balsara,Jamme2002} and experimentally \citep{Andreopoulos}.

The analytical studies focus mainly on the local mechanisms of turbulence amplification. However, when the finiteness of the flow becomes significant, non-local resonant effects can appear.

In our study we rely on the Linear Interaction Approximation which is mainly used since the pioneering work by \citet{Ribner54} to analyze the shock-turbulence interaction. Such an approximation seems to be reasonable if the turbulence is significantly subsonic and a low-amplitude one, which of course is the case under consideration.

To a certain extent, our analysis resembles an approach of \citep{Sari12}; however, it possesses greater versatility.

The outline of the paper is as follows: in Section 2 we describe the physical model of the shock wave produced by the point explosion, discuss its symmetrical (scaleless) properties, define similarity variables both for the unperturbed flow and perturbations, and pose a mathematical problem for the postshock perturbations with particular focus on derivation of the inner boundary conditions. Mathematically, the computational task is reduced to an eigenvalue problem for finding the amplitudes of the postshock perturbations if the amplitudes and frequency of outer perturbations are given. In Section 3 we present the results of numerical integration. We show that both for the vortex modes and for the entropy modes there exist conditions for infinite (in the linear approximation) resonant amplification. In Section 4 we discuss the domain of applicability of solutions found and present a brief summary.

\section{THE MODEL}

\subsection{Basic Equations and Unperturbed Flow}

All supernova remnants mentioned in Introduction are young or middle-aged objects with sizes $\sim 10$-$20$ pc for which approximation of adiabatic expansion holds.
Thus we base our consideration on an adiabatic shock wave model.

The magnetic field becomes dynamically important if magnetic pressure becomes comparable with dynamic and/or thermodynamic pressure. Simple estimates show that magnetic pressure becomes equal to the density of released supernova energy $10^{51}$ erg smeared out all over the spherical volume with a radius of 10 pc if the magnetic field attains the value $\sim 300 \mu$G. There are many magnetized remnants, and SN1006 with $\sim 170 \mu$G \citep{MC10} among them, for which B-field cannot be neglected. On the other hand, \citet{Warren04} estimate magnetic field value in SNR 0509-67.5 as 60 $\mu$G; therefore magnetic pressure is insignificant there at the adiabatic stage at least. In the present paper we limit ourselves to an examination of the pure hydrodynamical model neglecting the effects of magnetic pressure.

The shock wave dynamics is described by the system of gasdynamic equations for inviscid perfect gas with an adiabatic index $\gamma$
\begin{equation}\label{Continuity Equation}
\frac{\partial \rho}{\partial t}+{\rm div}(\rho\mathbf{v})=0\,,
\end{equation}
\begin{equation}\label{Euler Equation}
\frac{\partial (\rho\mathbf{v})}{\partial t}+{\rm div}(\rho\mathbf{v}\mathbf{v})+\nabla p=0,
\end{equation}
\begin{equation}\label{Energy Equation}
\frac{\partial}{\partial t}
\left(
\frac{\rho\mathbf{v}^2}{2}+ \frac{p}{(\gamma-1)}
\right) +
{\rm div}
\left(
\frac{\rho\mathbf{v}\mathbf{v}^2}{2}+ \frac{\gamma p \mathbf{v}}{(\gamma-1)}
\right)=0\,,
\end{equation}
which should be complemented by the outer boundary conditions at the strong shock front which in our case read as
\begin{equation}\label{mass}
-\rho_0 \mathbf{V}=\rho_s(\mathbf{v}_s-\mathbf{V})\,,
\end{equation}
\begin{equation}\label{impulse}
\rho_0 \mathbf{V}^2=\rho_s(\mathbf{v}_s-\mathbf{V})^2+p_s\,,
\end{equation}
\begin{equation}\label{energy}
\frac1 2 \mathbf{V}^2=
\frac1 2(\mathbf{v}_s-\mathbf{V})^2+\frac{\gamma}{\gamma-1}\frac{p_s}{\rho_s}\,.
\end{equation}
Here $\rho$ is the density, $\mathbf{v}$ the velocity, $p$ the pressure, and $\mathbf{V}$ is a shock front velocity.
Hereafter we denote by the subscripts ``0'', ``1''  the states outside and inside the shock-wave flow, respectively, and $``s"$ denotes the states just behind the shock front.

Further we consider both preshock and postshock flows as a superposition of the unperturbed (mean) flow and small perturbations
\begin{equation}\label{medium01}
\rho=\rho_{0,1}+ \delta\rho_{0,1}, \ \  \mathbf{v} =\mathbf{v}_{0,1}  + \delta\mathbf{v}_{0,1},\ \  p=p_{0,1} + \delta  p_{0,1}.
\end{equation}

The parameters of the mean ambient medium are considered as a motionless, zero-pressure state with spherically symmetric power-law density distribution
\begin{equation}\label{medium3}
 \rho_0=A/r^\omega, \qquad \mathbf{v}_0=0,\qquad  p_0=0,
\end{equation}
where $r$ is the spherical radius, $A$ and $\omega$ are certain constants.

We suppose that at the initial moment $t=0$ an instant release of energy $E_0$ takes place in the origin that generates a flow with a strong spherical shock wave.

The mean values inside the shock flow obey the self-similar Sedov solution \citep{Sedov59} in which the governing parameter, the similarity index  $\delta$ is equal to
\begin{equation}\label{delta}
\delta=\frac{2}{5-\omega}\,,
\end{equation}
characterizes the rate of the unperturbed shock front expansion,
\begin{equation}\label{R_s}
R_s(t)\sim t^\delta\,, \qquad  V = \frac{d R_s}{d t} = \delta \frac{R_s}{t}\,.
\end{equation}

Further we use variables for the mean flow
\begin{equation}\label{bezrazmer1}
\tilde{\rho}=\frac{\rho_1}{\rho_s}\equiv \tilde{\rho}(\xi), \ \  \tilde{v}=\frac{v_{r_1}}{v_{r_s}}\equiv \tilde{v}(\xi),  \ \
\tilde{p}= \frac {p_1}{p_s}\equiv \tilde{p}(\xi),
\end{equation}
 normalized by the postshock values
\begin{equation}\label{vprho_rs}
\rho_s=\rho_0\frac{\gamma+1}{\gamma-1},\quad  v_{r_s}=\frac{2V}{\gamma+1},\quad  p_s=\frac{2\rho_0 V^2}{\gamma+1},
\end{equation}
and the self-similar variable as the radial coordinate
\begin{equation}\label{xi_def}
 \xi=\frac{r}{R_s(t)}\,.
\end{equation}

In dimensionless variables \eqref{delta}-\eqref{xi_def} the system
 \eqref{Continuity Equation}--\eqref{Energy Equation} for the mean flow assumes the form
\begin{equation}\label{eq01}
\left(\frac{2}{\gamma+1}\tilde{v}-\xi\right) \frac{d \tilde{\rho}}{\tilde{\rho} d
\xi}+\frac{2}{\gamma+1}\frac{d \tilde{v}}{d
\xi}+\frac{4}{\gamma+1}\frac{\tilde{v}}{\xi}-\omega=0,
\end{equation}
\begin{equation}\label{eq02}
\left(\frac{2}{\gamma+1}\tilde{v}-\xi\right)\frac{d \tilde{v}}{d
\xi}+\frac{\gamma-1}{\gamma+1} \frac{d \tilde{p}}{\tilde{\rho}d \xi}
-\frac{3-\omega}{2}\tilde{v}=0,
\end{equation}
\begin{equation}\label{eq03}
\gamma\left(\frac{2}{\gamma+1}\tilde{v}-\xi\right) \frac{d\tilde{\rho}}{\rho d
\xi} - \left(\frac{2}{\gamma+1}\tilde{v}-\xi\right) \frac{d \tilde{p}}{\tilde{p} d
\xi} + \left(3-\gamma\omega \right)=0,
\end{equation}

One has two qualitatively different kinds of the unperturbed flow, the shell-like one with a hollow cavity inside ($\omega> \frac{7-\gamma}{\gamma+1}$) or the solid flow which extends to the center of symmetry ($\omega <  \frac{7-\gamma}{\gamma+1}$)  \citep{Sedov59}.
In physical applications the most important are the cases of either uniform background $\omega=0$ or the power-law decreasing density with the
slope $\omega >0$.  In what follows we assume $0\le \omega < 3$. The latter inequality guarantees mass integrability in the origin.

\subsection{Perturbations}

In describing perturbations we also use the nondimensionalized form:
\begin{equation}\label{bezrazmer}
\begin{split}
&f_{\rho_{0,1}}=\frac{\delta \rho_{0,1}}{\rho_s}, \ \ f_{v_{\xi_{0,1}}}=\frac{\delta v_{r_{0,1}}}{v_{r_s}},\ \  f_{v_{\tau_{0,1}}}=\frac{\delta v_{\tau_{0,1}}}{v_{r_s}},\\
&f_{v_{b_{0,1}}}=\frac{\delta v_{b_{0,1}}}{v_{r_s}}, \ \  f_{p_{0,1}}=\frac {\delta p_{0,1}}{p_s}.
\end{split}
\end{equation}
Here subscripts $\tau$ and $b$ stand for tangential and binormal components of velocity to be explained further and the subscript ``1'' for the postshock perturbations is further omitted.

The symmetry of the flow permits looking for a solution for perturbations as an expansion in the vector spherical harmonics. One can find different definitions of the vector spherical harmonics in the literature \citep{Barrera,Winch05}; we follow that of \citep{KovEremin}.

Heuristic arguments for this expansion are as follows.

The sphericity of the unperturbed flow prompts seeking the expansion of scalar functions in a form of factored out terms $\sim f(\xi) Y_{lm}(\theta,\varphi)$ known as solid spherical harmonics. Spherical functions $Y_{lm}(\theta,\varphi)$ compose a complete orthogonal basis in the space of functions defined on a sphere $S^2$.

The vector functions are expanded as follows.
The geometry of the problem defines three preferred directions. Two of them are determined through the gradient of solid spherical harmonics $\nabla \left( \xi^n Y_{lm}(\theta,\varphi) \right)$: the first, radial one, is given by the gradient of the radial part, $\mathbf{e}_{\xi}  = \nabla \xi$, the second direction is determined by the gradient of the spherical harmonic, $\nabla Y_{lm}(\theta,\varphi)$. By definition the latter one is tangential to the concentric sphere of radius $\xi$ and thus orthogonal to $\mathbf{e}_\xi$. The third direction is then determined unambiguously as an orthogonal complement.

The right-hand triple of unit vectors
\begin{equation}\label{exi}
\begin{split}
\mathbf{e}_\xi&=\nabla (\xi)\,,\\
\mathbf{e}_\tau&=\frac {\mathbf\nabla Y_{lm}(\theta,\varphi)}{|\mathbf\nabla Y_{lm}(\theta,\varphi)|}\,,\qquad l^2+m^2 \ne 0\,,\\
\mathbf{e}_b &= \frac{\mathbf{e}_\xi\times\nabla Y_{lm}(\theta,\varphi)}{|\mathbf{e}_\xi\times\nabla Y_{lm}(\theta,\varphi)|}\,.
\end{split}
\end{equation}
thus sets the `natural' orthogonal system of coordinates.

This new coordinate system allows splitting the nabla operator in expansions similar to \eqref{e1} to radial, $\nabla_\xi$, and tangential, $\nabla_\tau$, parts and to complement it by the binormal counterpart $\nabla_b$:
\begin{equation}
\begin{split}
{\nabla}_\xi = \mathbf{e}_\xi \frac{\partial}{\partial\xi}\,,  \quad
{\nabla}_\tau  &=
{\nabla}-{\nabla}_\xi = \frac{\mathbf{e}_\theta}{\xi}\frac{\partial}{\partial
\theta}+\frac{\mathbf{e}_\varphi}{\xi
\sin(\theta)}\frac{\partial}{\partial \varphi}\,,   \\
\nabla_b &=\mathbf{e}_\xi\times\mathbf{\nabla}
=\frac{\mathbf{e}_\varphi}{\xi}\frac{\partial}{\partial
\theta}-\frac{\mathbf{e}_\theta}{\xi \sin(\theta)}\frac{\partial}{\partial \varphi}\,.
\end{split}
\end{equation}
The natural system optimizes the analysis inasmuch as the binormal component of perturbed velocity degenerates and thus perturbations are reduced to two-dimensional ones.

Ultimately, the symmetry and similarity of the flow allow us to look for a solution for perturbations as an expansion in the
solid spherical harmonics
\begin{equation}\label{e1}
\begin{split}
&f_{\rho}(\boldsymbol{\xi},t) =\sum\limits_{l=0}^\infty\sum\limits_{m=-l}^l
f_{\rho_{lm}}(\xi) Y_{lm}(\theta,\varphi) t^{-is_{l}},\\
&\mathbf{f}_{v}(\boldsymbol{\xi},t) = \sum\limits_{l=0}^\infty\sum\limits_{m=-l}^l
\left(       f_{v_{\xi_{lm}}}(\xi) Y_{lm}(\theta,\varphi)\mathbf{e}_\xi
\right. \\ & \left.
   +   f_{v_{\tau_{lm}}}(\xi) \xi \mathbf{\nabla}_\tau Y_{lm}(\theta,\varphi)
   +   f_{v_{b_{lm}}}(\xi) \xi \nabla_b Y_{lm}(\theta,\varphi)  \right) t^{-is_{l}},
  \\
&f_{p}(\boldsymbol{\xi},t)=\sum\limits_{l=0}^\infty\sum\limits_{m=-l}^l
f_{p_{lm}}(\xi) Y_{lm}(\theta,\varphi) t^{-is_{l}}.
\end{split}
\end{equation}
Here $\boldsymbol{\xi} = (\xi,\theta, \varphi )$ has a meaning of the radius vector in a comoving spherical coordinate system, $s_{l}$ is a frequency in the `comoving' logarithmic time scale $\tilde{t}=\log (t/t_0)$ and the frequency does not depend on $m$ since the differential equations for perturbations do not depend on the $m$-wavenumber.

The radial dependencies of perturbations $f_i(\xi)$ are found by solving the system of linearized hydrodynamic equations which can be obtained from the system \eqref{Continuity Equation}-\eqref{Energy Equation} with the help of \eqref{bezrazmer1}-\eqref{xi_def},\eqref{bezrazmer},\eqref{e1}. Orthogonality of harmonics allows  uncoupling and factorization of equations \citep{RV91}:
\begin{multline}\label{eqv1}
\left(\tilde{v}-\frac{\gamma+1}{2}\xi\right)\frac{d f_\rho}{d \xi} +\tilde{\rho}\frac{d f_{v_\xi}}{d \xi} \\
+\left(\frac{d \tilde{v}}{d \xi}+2\frac{\tilde{v}}{\xi} - \frac{\gamma+1}{2}\omega -\frac 1 4
(\gamma+1)(5-\omega)is\right)f_\rho \\
+\left(\frac{d \tilde{\rho}}{d \xi}+2\frac {\tilde{\rho}} {\xi} \right)f_{v_\xi} - l(l+1)\frac{\tilde{\rho}}{\xi}f_{v_\tau}=0\,,
\end{multline}
\begin{multline}\label{eqv2}
\left(\tilde{v}-\frac{\gamma+1}{2}\xi\right)\tilde{\rho}\frac{d f_{v_\xi}}{d
\xi}+\frac{\gamma-1}{2}\frac{d f_p}{d \xi}-\frac{\gamma-1}{2\tilde{\rho}}\frac{d
\tilde{p}}{d \xi}f_\rho \\
+ \left(\frac{d \tilde{v}}{d \xi}-\frac 1 4 (\gamma+1)(3-\omega) -\frac 1 4 (\gamma+1)(5-\omega)is\right)\tilde{\rho}
f_{v_\xi}\\
=0\,,
\end{multline}
\begin{multline}\label{eqv3}
\left(\tilde{v}-\frac{\gamma+1}{2}\xi\right)\tilde{\rho}\frac{d f_{v_\tau}}{d
\xi}\\
+\left(\frac{d \tilde{v}}{d \xi}-\frac 1 4 (\gamma+1)(3-\omega) -\frac 1 4 (\gamma+1)(5-\omega)is\right)\tilde{\rho}
f_{v_{\tau}}\\
+\frac{\gamma-1}{2\xi}f_p=0\,,
\end{multline}
\begin{multline}\label{eqv4}
\left(\tilde{v}-\frac{\gamma+1}{2}\xi\right)\frac{d f_{v_b}}{d \xi}\\
+\left(\frac{d
\tilde{v}}{d \xi}-\frac 1 4 (\gamma+1)(3-\omega)
-\frac 1 4 (\gamma+1)(5-\omega)is\right) f_{v_{b}}\\
=0\,,
\end{multline}
\begin{multline}\label{eqv5}
-\gamma\left(\tilde{v}-\frac{\gamma+1}{2}\xi\right)\frac{d f_\rho}{d \xi}
+\left(\tilde{v}-\frac{\gamma+1}{2}\xi\right)\frac{\tilde{\rho}}{\tilde p}\frac{d
f_p}{d \xi}\\
+\left(\frac{1}{\tilde{p}}\frac{d{\tilde{p}}}{d
\xi}-\frac{\gamma}{\tilde{\rho}}\frac{d \tilde{\rho}}{d \xi}
\right)\tilde{\rho}f_{v_\xi}\\
+\gamma\left(\left(\tilde{v}-\frac{\gamma+1}{2}\xi\right)
\frac{1}{\tilde{\rho}}\frac{d \tilde{\rho}}{d \xi}+\frac 1 4
(\gamma+1)(5-\omega)is\right)
f_{\rho} \\
-\left(\left(\tilde{v}-\frac{\gamma+1}{2}\xi\right)
\frac{1}{\tilde{p}}\frac{d\tilde{p}}{d\xi}+\frac 1 4
(\gamma+1)(5-\omega)is\right)\frac{\tilde{\rho}}{\tilde{p}} f_{p}\\
=0\,.
\end{multline}
Hereinafter the subscripts $l$ and $m$ are omitted.
The typo in the bracketed expression at the first term in \eqref{eqv5} from \citep{RV91} is corrected as well.

It is worth noting once more that the system \eqref{eqv1}-\eqref{eqv5} is $m$-wavenumber degenerated and that the variable $f_{v_b}$ uncouples (Eq.~\eqref{eqv4}).

\subsection{External Perturbations}

The system \eqref{eqv1}-\eqref{eqv5} describes perturbations both inside and outside the shock wave.
First consider the external to the shock perturbations coded by the subscript ``0''.

The radial function $f(\xi)$  in solid harmonics is usually chosen as an expansion in eigenfunctions of some operator relevant to the problem \citep{Landafshitz}.
For example, for the Laplace equation
\begin{displaymath}
{\Delta}\phi=0
\end{displaymath}
the radial parts are just power functions; the expression
\begin{displaymath}
 \phi =  \sum\limits_{n,l=0}^\infty\sum\limits_{m=-l}^l  \phi_{lmn} \xi^n Y_{lm}(\theta,\varphi)
\end{displaymath}
yields a general solution of the Laplace equation in a ball.

This power-law expansion of the radial function is nothing but the Taylor series.
Considering the perturbations
as a regular scalar or vector field bounded within the finite domain we use the power-law solid spherical harmonics
\begin{equation}
 f_{i_0}(\xi) =  \sum\limits_{n=0}^\infty f_{i_{0n}} \xi^n, \quad i=(\rho, v_{\xi}, v_{\tau}, v_{b},p)
\end{equation}
as a basis for the external perturbations expansion.

For simplicity, we further limit ourselves to the principal mode $n=0$, the subscript $n$ will, therefore, be omitted.
It follows therefrom that the absolute external perturbations $\delta\rho_0$, $\delta \mathbf{v}_0$, $\delta p_0$ under consideration are spatially uniform.

Assuming the turbulent motion in the ambient medium to be a slow and subsonic one, and the outside pressure perturbations negligible compared with the pressure behind the strong shock wave, we can neglect the external acoustic perturbations ($f_{p0}\to 0$) and consider just two different modes, either (i) the vortical mode ($\mathbf{f}_{v0}\neq0,\ f_{\rho 0} =  f_{p0} = 0$), or (ii) entropy perturbations ($f_{\rho 0}\neq0,\ \mathbf{f}_{v0} = f_{p0} = 0$). In any case, the acoustic perturbations ($f_{p0}\neq 0$) cannot be considered self-consistently as far as they violate the similarity condition.

As we know, the vortex and entropy modes are frozen in the environment which we regard as motionless, thus they have zero frequency in the rigid reference frame. At the same time, relative perturbations defined in the comoving frame and normalized according to \eqref{bezrazmer} should, of course, be time-dependent.

Stationary preshock perturbations must obey the linearized continuity equation
\begin{equation}\label{selfper1}
{\rm div}(\rho_0 \delta \mathbf{v}_0) =0\,.
\end{equation}
Taking into account expansion in spherical harmonics, with the aid of \eqref{medium3} we get from
\eqref{selfper1} the relation between the radial and tangential velocity components
\begin{equation}\label{svas}
\delta v_{\xi_{0}}=\frac{l(l+1)}{2-\omega}\delta v_{\tau_{0}}.
\end{equation}
This is a unique constraint for the outside perturbations: the other linearized hydrodynamics equations for the momentum and energy conservation are satisfied automatically.

The binormal component of velocity does not enter the equation \eqref{selfper1} and can be nullified so long as it does not couple with other terms due to neither boundary conditions on the shock front nor hydrodynamic equations.

In total, one of three perturbations $\delta v_{r_{0lm}}$, $\delta v_{\tau_{0lm}}$, $\delta \rho_{0_{lm}}$ can be considered free whereas the other two should be dependent ones.
For the entropy mode we assume
\begin{equation}\label{entropy_mode}
f_{\rho_{0}}\ne 0\,, \qquad f_{v_{\xi_{0}}}=f_{v_{\tau_{0}}}=0\,,
\end{equation}
for the vortical mode we have
\begin{equation}\label{svasf3}
f_{\rho_{0}}= 0\,, \quad f_{v_{\xi_{0}}}= \frac{l(l+1)}{2-\omega} f_{v_{\tau_{0}}}, \quad f_{v_{\tau_{0}}}\ne 0\,.
\end{equation}

By virtue of stationarity of absolute perturbation, $\delta\rho_0\propto t^0$, and self-similarity, $\rho_s\propto t^{-\delta\omega}$, (see equations \eqref{medium3}, \eqref{R_s})
for the relative perturbation we get
\begin{equation}\label{statrel1}
f_{\rho_{0_{}}}=\frac{\delta\rho_{0_{}}}{\rho_s}=\frac{\gamma-1}{\gamma+1}\frac{\delta\rho_{0_{}}}{\rho_0}\propto
t^{\delta\omega}\,.
\end{equation}
With regard to proportionality $f_{\rho_{0}}\propto t^{-is}$ according to \eqref{e1} and using the definition \eqref{delta}, we finally find the frequency for the entropy mode
\begin{equation}\label{statrel2}
s^{(e)}=\frac{2\omega}{5-\omega}i\,.
\end{equation}
For the vortex mode with the aid of relations $\delta \mathbf{v}_0\propto t^0$, $v_s\propto t^{\delta-1}$ we similarly get
\begin{equation}\label{statrel3}
s^{(v)}=\frac{3-\omega}{5-\omega}i\,.
\end{equation}
The frequencies \eqref{statrel2},\eqref{statrel3} can equally well be found upon substituting \eqref{entropy_mode}, \eqref{svasf3} into \eqref{eqv1}-\eqref{eqv5}.

\subsection{Perturbations Inside the Shock Wave}

The radial dependencies $f_i(\xi)$  of the solution for the perturbations inside the shock wave
are not presumed to be power series expansion; rather, they are to be found by solving the system \eqref{eqv1}-\eqref{eqv5} with the inner and outer boundary conditions.

\subsection{Boundary Conditions for Perturbations}

\subsubsection{Outer Boundary Conditions}

Linearized boundary conditions for the amplitudes of perturbations can be deduced from the general boundary conditions \eqref{mass}-\eqref{energy} with regard to the fact of the shock surface warping:
\begin{equation}\label{Renken7}
f_{\rho_s} = -\left(\omega +\frac{d \tilde{\rho}}{d \xi}\bigg\vert_s \right)\eta +\frac{\gamma+1}{\gamma-1}f_{\rho_0},
\end{equation}
\begin{equation}\label{Renken6}
\begin{split}
f_{p_s}= &\left(2-\omega - (5-\omega) i s - \frac{d \tilde{p}}{d \xi}\bigg\vert_s \right)\eta \\
         &+\frac{\gamma+1}{\gamma-1}f_{\rho_0}-\frac{4}{\gamma+1}f_{v_{\xi_0}},
\end{split}
\end{equation}
\begin{equation}\label{Renken8}
f_{v_{\xi_s}} = \left( 1-\frac{(5-\omega)}{2}is-\frac{d \tilde{v}}{d \xi}\bigg\vert_s \right)\eta +\frac{\gamma-1}{\gamma+1}f_{v_{\xi_0}},
\end{equation}
\begin{equation}\label{Renken9}
f_{v_{\tau_s}} = -\eta +f_{v_{\tau_0}},
\end{equation}
\begin{equation}\label{Renken10}
f_{v_{b_s}} =f_{v_{b_0}} =0.
\end{equation}
Corrugation of the shock front is described here by the small dimensionless parameter
\begin{equation}\label{eta}
\eta(\theta,\varphi,t) \equiv \frac{\Delta R_s}{R_s}=\sum_{l,m}\eta_{lm} Y_{lm}(\theta,\varphi) t^{-is_l} \ll 1\,,
\end{equation}
where $\Delta R_s(\theta,\varphi,t)$ is a small shift of the front from its mean value $R_s(t)$.

\subsubsection{Inner Boundary Condition at the Center of Symmetry}

The relationship between characteristics requires only one inner boundary condition.

It is notable that derivation of the correct inner boundary condition has a long history. \citet{RV87} derived the equation $\delta p(0)=0$ which was criticized by \citet{Kushnir}.
Indeed, the condition of null perturbed pressure seems to go beyond the bounds of common sense. If we uniformly compress an element of fluid, pressure will definitely grow inside the fluid. \citet{Kushnir} proposed another condition which they formulated in an asymptotic form suggesting a lack of similarity at the origin due to specific initial conditions (non-point and probably non-instant energy release). This condition being adapted for numerical calculation, no analytical expression was presented.

In fact the condition at the inner boundary, as well as the outer boundary, is actually a {\it no-source condition} and can be explicitly derived from Equations \eqref{Continuity Equation}-\eqref{Energy Equation} expressing conservation laws.

Let us integrate the linearized divergent terms from the Equations \eqref{Continuity Equation}-\eqref{Energy Equation} over the small sphere $S_{\epsilon}$ of vanishingly small radius $\epsilon$ and the volume $V_{\epsilon}$ centered at the origin.

For the linearized  mass flux we get
\begin{equation}\label{InnerBC_mass}
\begin{split}
\delta I_{mass} &=  \int\limits_{V_{\epsilon}} {\rm div} \left[\delta (\rho \mathbf{v})\right] dV =  \oint\limits_{S_{\epsilon}} (\delta \rho \mathbf{v}_0 + \rho_0 \delta \mathbf{v}) d\mathbf{S} \\
&=  (\delta \rho(\epsilon) v_0(\epsilon) + \rho_0(\epsilon) \delta v_{\xi}(\epsilon)) \epsilon^2 \oint\limits_{4\pi} {Y_{lm} {\rm d}\Omega}\,,
\end{split}
\end{equation}
where $\Omega$ stands here for the solid angle.

In a similar manner for the fluxes of momentum and energy we have
\begin{equation}\label{InnerBC_mom}
\begin{split}
&\delta \mathbf{I}_{mom} = \int\limits_{V_{\epsilon}} {\rm div} \left[\delta \left(\rho \mathbf{v}\mathbf{v} +p \hat{I}\right)\right] dV \\
= &\left(\delta \rho(\epsilon) v^2_0(\epsilon) + 2 \rho_0(\epsilon) v_0(\epsilon) \delta v_{\xi}(\epsilon) + \delta p(\epsilon)\right)\epsilon^2 \int\limits_{4\pi} Y_{lm}\mathbf{e}_\xi {\rm d}\Omega \\
+ &      \rho_0(\epsilon) v_0(\epsilon) \delta v_{\tau}(\epsilon)\epsilon^3 \int\limits_{4\pi} \mathbf{\nabla}_\tau Y_{lm} {\rm d}\Omega\,,
\end{split}
\end{equation}
\begin{equation}\label{InnerBC_ener}
\begin{split}
&\delta I_{ener} = \int\limits_{V_{\epsilon}} {\rm div} \left[\delta \left(\rho \mathbf{v} \frac{v^2}{2}  + \frac{\gamma p \mathbf{v}}{\gamma-1} \right)\right] dV \\
= &\frac{1}{2} \left[ \delta \rho(\epsilon) v^3_0(\epsilon) + 3 \rho_0(\epsilon) v^2_0(\epsilon) \delta v_{\xi}(\epsilon) \right] \epsilon^2  \int\limits_{4\pi} Y_{lm} {\rm d}\Omega \\
+ & \frac{\gamma  }{\gamma-1} \left[\delta  p(\epsilon) v_0(\epsilon)  + \delta v_{\xi}(\epsilon) p_0(\epsilon) \right] \epsilon^2  \int\limits_{4\pi} Y_{lm} {\rm d}\Omega\,.
\end{split}
\end{equation}

The vector integral in Equation \eqref{InnerBC_mom} $\int\limits_{4\pi} Y_{lm}\mathbf{e}_\xi {\rm d}\Omega$ is nonzero only for the dipole $l=1$, the scalar integral $\int\limits_{4\pi} Y_{lm} {\rm d}\Omega$ in Equation \eqref{InnerBC_mass} and \eqref{InnerBC_ener} becomes zero in all instances except for the monopole $l=0$. Although the fluxes integrated over full solid angle can vanish identically due to the specific nature of the spherical harmonics, the sectorial flux density may not.

As an inner boundary condition we demand that the flux must vanish in each direction, that is,
\begin{equation}\label{InnerBC_mass_sector}
  \lim_{\xi\to 0} \left[f_{\rho}(\xi) \tilde{v}(\xi) + \tilde{\rho}(\xi) f_{v_{\xi}}(\xi)\right] \xi^2 =0 ,
\end{equation}
\begin{equation}\label{InnerBC_mom_sector_xi}
\lim_{\xi\to 0}
\left[f_{\rho}(\xi) \tilde{v}^2(\xi) + 2 \tilde{\rho}(\xi) \tilde{v}(\xi) f_{v_{\xi}}(\xi) + f_p(\xi)\right]\xi^2 =0  ,
\end{equation}
\begin{equation}\label{InnerBC_mom_sector_tau}
\lim_{\xi\to 0}  \tilde{\rho}(\xi) \tilde{v}(\xi) f_{v_{\tau}}(\xi)\xi^2 =0  ,
\end{equation}
\begin{equation}\label{InnerBC_ener_sector}
\begin{split}
\lim_{\xi\to 0}
   & \left[ \frac{1}{2} f_{\rho}(\xi) \tilde{v}^3(\xi) + \frac{3}{2}  \tilde{\rho}(\xi) \tilde{v}^2(\xi) f_{v_{\xi}}(\xi) \right. \\
+ &  \left.  \frac{\gamma  }{\gamma-1} \left(f_p(\xi) \tilde{v}(\xi)  + f_{v_{\xi}}(\xi) \tilde{p}(\xi) \right) \right] \xi^2 =0 .
\end{split}
\end{equation}

The very unique inner boundary condition is the one majorizing term in Equations \eqref{InnerBC_mass_sector}-\eqref{InnerBC_ener_sector} that guarantees the fulfillment of all four conditions. To find this majorizing term we have to determine the relation among perturbed variables.

First, we note that the unperturbed variables asymptotically behave at the origin  $(\xi\to 0)$ as \citep{Sedov59}
\begin{equation}\label{Sedov_asympt}
\begin{split}
&\rho_0 = a_0\xi^{a-2}, \qquad v_0 = b_0\xi, \qquad p_0 = c_0 +c_1\xi^a,  \\
& a ={1+\gamma(2-\omega)\over \gamma-1}, \\
&\kappa_1 = {2(3-\omega)\over (\omega-5)(\gamma-1)}, \qquad \kappa_2 = { (1+\gamma)\omega-6\over 6-3\gamma-\omega }, \\
&\kappa_3 = { \gamma^2(\omega^2-4\omega+13)+ \gamma(\omega+1)(\omega-7) -2\omega+12 \over (\omega-5)(3\gamma+\omega-6)(3\gamma-1) }, \\
&a_0 = \left({1+\gamma\over 2\gamma}\right)^{\kappa_1} \left({1+\gamma \over \gamma}\right)^{\kappa_2}
\left({\omega-{2\gamma+1\over \gamma}\over \omega-{7-\gamma\over 1+\gamma}}\right)^{{3-\omega\over \gamma-1}\kappa_3},\\
&c_0 = \left({1+\gamma\over 2\gamma}\right)^{6\over 5-\omega} \left({1+\gamma \over \gamma}\right)^{\kappa_2+1}
\left({\omega-{2\gamma+1\over \gamma}\over \omega-{7-\gamma\over 1+\gamma}}\right)^{3\kappa_3},\\
&b_0 = {\gamma+1\over 2\gamma}, \qquad c_1 = a_0 b_0^2 {\gamma (5-\omega) -2 \over \gamma(2-\omega)+1}.
\end{split}
\end{equation}
Here $a>1$ for $0\le \omega <(7-\gamma)/(\gamma+1)$ and $\gamma>1$. This particularly means that the term $d\tilde{p}/\tilde{p}d\xi$ is asymptotically negligible compared to other terms in Equation \eqref{eq03} as well as in Equation \eqref{eqv5}
but the other term with pressure $\sim \tilde{\rho}/\tilde{p}$ and the pressure gradient itself matter.

The coefficients of the system \eqref{eqv1}-\eqref{eqv3}, \eqref{eqv5} asymptotically behave as power functions at the origin. Therefor,e we look for solutions for perturbations as power laws in $\xi$. From Equation \eqref{eqv1} we find $f_{v_{\xi}} \sim \xi^{3-a}f_{\rho}$,  $f_{v_{\tau}} \sim \xi^{3-a}f_{\rho}$; from Equation \eqref{eqv2} we get also $f_p\sim \xi^{a-1} f_{v_{\xi}} \sim \xi^{2}f_{\rho}$. Equations \eqref{eqv3} and \eqref{eqv5} confirm relations found.

Suppose
\begin{equation}\label{frho_alpha}
f_{\rho} = F_{\rho 0} \xi^{\alpha}
\end{equation}
with some constant amplitude $f_{\rho 0}$, then, similarly, we get
\begin{equation}\label{fv_fp_alpha}
f_{v_{\xi}} = F_{v_{\xi} 0} \xi^{3-a+\alpha}, \ \  f_{v_{\tau}} = F_{v_{\tau} 0} \xi^{3-a+\alpha}, \ \ f_{p} = F_{p 0} \xi^{2+\alpha}.
\end{equation}
The solution for perturbations then can be assembled in a vector form as
\begin{equation}\label{f_f0xi}
\mathbf{f} =\mathbf{F}_0(\xi)\xi^{\alpha}, \ \ \mathbf{F}_0(\xi) = (F_{\rho 0}, F_{v_{\xi} 0}\xi^{3-a}, F_{v_{\tau} 0}\xi^{3-a}, F_{p 0}\xi^{2})^T.
\end{equation}
Substituting Equations \eqref{frho_alpha}, \eqref{fv_fp_alpha} into Equations \eqref{eqv1}-\eqref{eqv3}, \eqref{eqv5} we get the system of homogeneous linear algebraic equations for the amplitudes $(F_{\rho 0}, F_{v_{\xi} 0}, F_{v_{\tau} 0}, F_{p 0})$, from which we find four roots for $\alpha$ as the eigenvalues.
The asymptotic solution is ultimately a superposition of four branches with some amplitudes $C_i$
\begin{equation}\label{f_sol}
\mathbf{f} = \sum_{i=1}^{4} C_i \mathbf{F}_{0i}(\xi) \xi^{\alpha_i}.
\end{equation}
One of the root, say $\alpha_1$, describes a divergent solution for a point source while three other are the regular ones.
The analytical expression for this exponent $\alpha_1$ is quite cumbersome; its existence can be checked numerically. However we manage without its explicit expression by specifying conditions directly to amplitudes.

Substituting Equations \eqref{frho_alpha}, \eqref{fv_fp_alpha} to the boundary fluxes \eqref{InnerBC_mass_sector}-\eqref{InnerBC_ener_sector}, we find the asymptotics $\xi^{3+\alpha}$  for the mass flux and all terms in \eqref{InnerBC_mass_sector} are of the same order. The asymptotics for the momentum fluxes is $\xi^{4+\alpha}$ and all terms in \eqref{InnerBC_mom_sector_xi}-\eqref{InnerBC_mom_sector_tau} are of the same order again. And finally for the energy flux the asymptotics is $\xi^{5-a+\alpha}$ which is determined  by the dominating term $f_{v_{\xi}}\xi^2$. The majorizing term among these is that whose power exponent is the minimum one. If $a<2$, the mass flux dominates, if $a>2$ the energy flux does. In the latter case the majorizing term is $f_{v_{\xi}}\xi^2$, in the former case vanishing of $f_{v_{\xi}}\xi^2$ implies vanishing of all other perturbation terms.

Thus, the inner boundary no-source condition yields an equation
\begin{equation}\label{obolochkainer}
f_{v_{\xi}}\xi^2\Big\vert_{\xi=0} = 0.
\end{equation}

\subsubsection{Inner Boundary Condition for the Flow with Hollow}

Inner boundary condition for the shell-like flow can be derived as the condition for a contact interface in the point $\xi_{in}>0$ separating gas and vacuum. Two conditions must be fulfilled simultaneously for the contact discontinuity: (i) there is no flux through the shifted boundary and (ii) the pressure at the shifted boundary is identically zero. Equations for these can be practically derived from the general matching conditions \eqref{mass}-\eqref{energy} by nulling l.h.s. and by substituting the displacement of the inner boundary $\eta_{in}$ for $\eta$ and the subscript ``in'' for ``s''. We obtain two conditions for the velocity and for the pressure \citep{RV91}
\begin{equation}\label{Inner0v}
f_{v_{\xi}} = -\left( \frac{\gamma+1}{2}+ \frac{is}{4} (\gamma+1)(5-\omega)+\frac{d \tilde{v}_{\xi}}{d \xi}\bigg\vert_{\xi_{in}}\right)\eta_{in},
\end{equation}
\begin{equation}\label{Inner0p}
f_{p} = -\frac{d \tilde{p}}{d \xi}\bigg\vert_{\xi_{in}}\eta_{in}.
\end{equation}
Accurate within a misprint these formulae coincide with those deduced by \citet{RV91}.
Elimination of $\eta_{in}$ sets seemingly desired relation between two unknown quantities.

A more thorough analysis shows however that while $(d \tilde{v}_{\xi}/d \xi)\vert_{\xi_{in}}$ in Equation \eqref{Inner0v} remains finite and nonzero, the derivative $(d \tilde{p}/d \xi)\vert_{\xi_{in}}$ in Equation \eqref{Inner0p} can either diverge or vanish. Indeed, we know from \cite{Sedov59} an asymptotical behavior of unperturbed pressure near the inner boundary
\begin{equation}\label{Inner3}
 \tilde{p}(\xi)
 \approx \tilde{p}_{0}\cdot(\xi-\xi_{in})^{\mu +1} , \qquad \tilde{p}_{0}={\rm const}\,,
\end{equation}
where $\mu$ determined as
\begin{equation}\label{Inner4}
\mu=\frac{\omega(1+\gamma)-6}{6-3\gamma-\omega}\,, \qquad  \frac{7-\gamma}{\gamma+1}\le \omega < 3\,,
\end{equation}
reverses sign in the tolerance interval of $\omega$. Thus, Equations \eqref{Inner0v}-\eqref{Inner0p} in reality are a singular system.

The correct condition can be derived by exceeding the limits of linear approximation and by using the property (ii) of the contact discontinuity.
Since from now on, we temporally use non-linear dependencies, let us temporally consider all perturbations as real variables.
For the perturbed pressure  $\tilde{p}_1^{\prime}$ at the perturbed inner boundary in the first order we have
\begin{equation}\label{Inner1}
\begin{split}
0 &= \tilde{p}^{\prime}(\xi_{in}+\eta_{in}) = \tilde{p}_{0}\cdot(\eta_{in})^{\mu +1} + f_p\,, \ \ \ \eta_{in}>0\,; \\
0 &=f_p\,, \qquad\qquad\qquad\qquad\qquad\qquad\qquad\   \eta_{in}<0.
\end{split}
\end{equation}
Taking into account Equations \eqref{Inner0v}, \eqref{Inner0p}, \eqref{Inner3}, \eqref{Inner1}, we finally get a non-linear relation:
\begin{equation}\label{Inner2}
f_p \propto f_{v_{\xi}}^{\mu +1}\,, \quad \eta_{in}>0\,.
\end{equation}

In the case $\mu>0$, considering the vanishing amplitude limit,  $f_{v_{\xi}}\to 0$, we find $|f_p|\ll |f_{v_{\xi}}|$ or asymptotically
\begin{equation}\label{In1bez}
f_ {p}\Big|_{\xi_{in}}=0\,.
\end{equation}
For $\mu<0$ the situation is reversed
\begin{equation}\label{In2bez}
f_{v_{\xi}}\Big|_{\xi_{in}}=0\,.
\end{equation}
In the particular case $\mu=0$ we come back to the initial Equations \eqref{Inner0v}-\eqref{Inner0p}.

\section{AMPLIFICATION OF DISTURBANCES BY THE SHOCK WAVE}

In order to know whether the shock wave amplifies perturbations or not, we have to find the structure of the perturbed postshock flow.
We expect amplification first of all in the case of resonance between the free (i.e. self-excited) oscillations of the flow and oscillations excited by the external perturbations if resonance only exists.

\begin{figure}
\epsscale{1.17}
\plotone{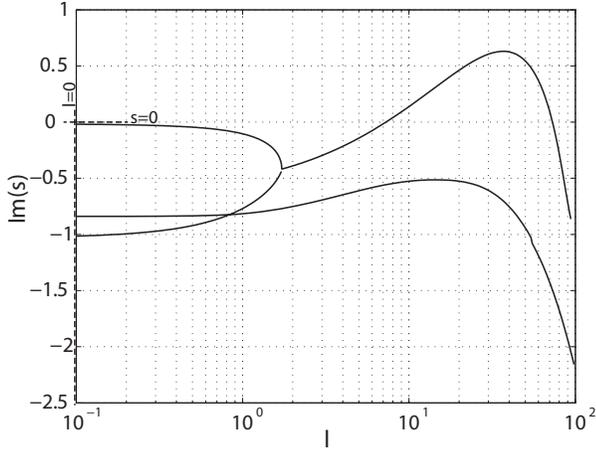}
\caption{
The image part of dispersion curves (frequency $s$ vs orbital wavenumber $l$)
for the blast wave free oscillations in the case of the uniform background ($\omega=0$, $\gamma=1.1$.
Positive ${\rm Im}(s)$ correspond to unstable oscillations. Dashed lines indicate location of the resonance point.}\label{spectrg11om0}
\end{figure}

\begin{figure}
\epsscale{1.17}
\plotone{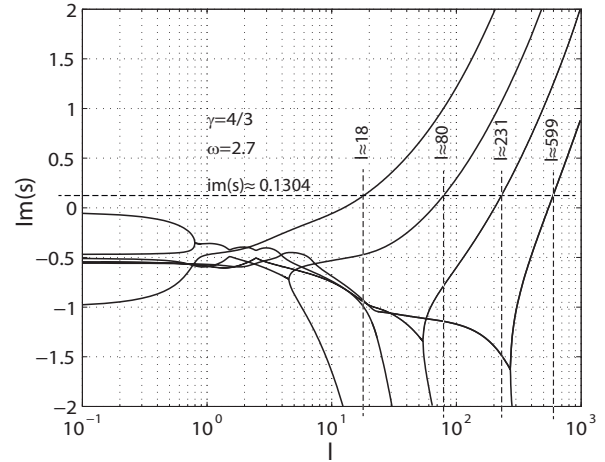}\caption{The same as in Fig.~\ref{spectrg11om0} but for the nonuniform background ($\omega=2.7$, $\gamma=4/3$).}\label{spectrg43om27}
\end{figure}

\subsection{Self-excited Oscillations and Resonance Conditions}

The eigenmodes of the shock flow were studied in detail by \citet{RV87,RV88,RV89}, \citet{Kushnir} for the filled flow and by \citet{Goodman2} and \citet{RV91} for the shell-like flow.
To find the resonance conditions we calculated first the dispersion relations for free oscillations guided by the approach of the aforementioned authors.

Zeroizing the amplitudes of the external perturbations, one finds the structure of the perturbed flow inside the shock wave by solving in the interval $\xi_{in}\le\xi\le 1$, $\xi_{in}\ge 0$ the system of linear ODEs \eqref{eqv1}-\eqref{eqv3},\eqref{eqv5} for unknown variables $f_{v_{\rho}}(\xi)$, $f_{v_{\xi}}(\xi)$, $f_{v_{\tau}}(\xi)$, $f_ {p}(\xi)$, $\eta$ with the outer \eqref{Renken6}-\eqref{Renken9} and inner \eqref{obolochkainer}, \eqref{In1bez} or \eqref{In2bez} boundary conditions. The frequency $s$ of free oscillations is then found as an eigenvalue.

The typical dispersion dependencies of $s$ on the wave number $l$ for the initial few lower order modes
are presented in Figures.~\ref{spectrg11om0} (solid flow, $\omega=0$,  $\gamma=1.1$) and \ref{spectrg43om27} (shell flow, $\omega=2.7$,  $\gamma=4/3$).
Each mode meets a particular number of nodes of oscillations along radius. The upper mode in Figures.~\ref{spectrg11om0} and \ref{spectrg43om27} is the principal mode (no nodes), the other ones, from top downward, higher order, so-called reflection modes have 1, 2 and so on indefinitely nodes.
The flow with $\gamma$ close to 1 approximates the radiative flow to some degree, while $\gamma$ close to 5/3 pertains to pure adiabatic flow.

Suppose the external perturbation has frequency $s^*$ and the wavenumber $l^*$. It will be the resonant one if the point $(s^*,l^*)$ is disposed exactly on the dispersion curve $s(l)$ (represented by dashed lines in Figures.~\ref{spectrg11om0} and \ref{spectrg43om27}) or in its close vicinity. We would like to remind that $s^*$ and $s(l)$ take on the complex values in the general case and hence $(s^*,l^*)$ and $s(l)$ must coincide not only in the imaginary parts as is displayed in Figures.~\ref{spectrg11om0}-\ref{spectrg43om27} but also in real parts.
Since we have adopted the uniform distribution of external perturbations, their frequencies $s^*$ can only be imaginary ones according to Equations \eqref{statrel2}, \eqref{statrel3} and  non-negative if $\omega\ge 0$. Then the equality of frequencies is possible in the increments domain where ${\rm Im} s(l)\ge 0$. Numerical linear analysis \citep{RV87,RV91} shows that the unstable regimes exist at $\gamma < 1.2$ for the solid flow ($\omega=0$) and for any permitted $\gamma$ in the case of the shell flow.

The condition for resonance
\begin{equation}\label{Reson_condition}
{\rm Re}\, s(l^*)=0, \quad {\rm Im}\, s(l^*)=s^*\ge 0
\end{equation}
is fulfilled for the entropy mode in the case of the solid flow only in the origin at the point
\begin{equation}\label{Reson_point_entropy_om0}
s^*=0, \quad l^*=0; \qquad \omega=0\,,
\end{equation}
whereas the unstable section $10^1 \lesssim l \lesssim 10^2$ (Figure ~\ref{spectrg11om0}) corresponds to nonzero ${\rm Re}\,s(l)$.

In the shell case the resonance frequency for the entropy mode cuts the sequence of points with high wave numbers
\begin{equation}\label{Reson_point_entropy_om27}
s^*=i\cdot  2.347...\,,\quad  l^*\approx 230,\  700, ...\,; \qquad \omega=2.7\,;
\end{equation}
they are located beyond the object area in Figure ~\ref{spectrg43om27}.

The resonance condition for the vortex mode
\begin{equation}\label{Reson_point_vortex_om0}
s^*=0.6 i\,; \qquad \omega=0\,,
\end{equation}
is never fulfilled for the solid flow by the same token  ${\rm Re}\,s(l)\ne 0$ but is fulfilled at growing unstable branches for the shell flow and specifies another sequence of points  (Figure ~\ref{spectrg43om27})
\begin{equation}\label{Reson_point_vortex_om27}
s^*=i\cdot 0.1304...\,, \quad l^*\approx 18, ...\, ; \qquad \omega=2.7\,.
\end{equation}

\subsection{Forced Oscillations. Amplification of Disturbances behind the Shock Front}

In order to find the structure of the postshock flow perturbed by external disturbances, we have to find the eigenfunctions considering the arising mathematical problem as an eigenvalue problem.

\begin{figure}
\epsscale{1.17}
\plotone{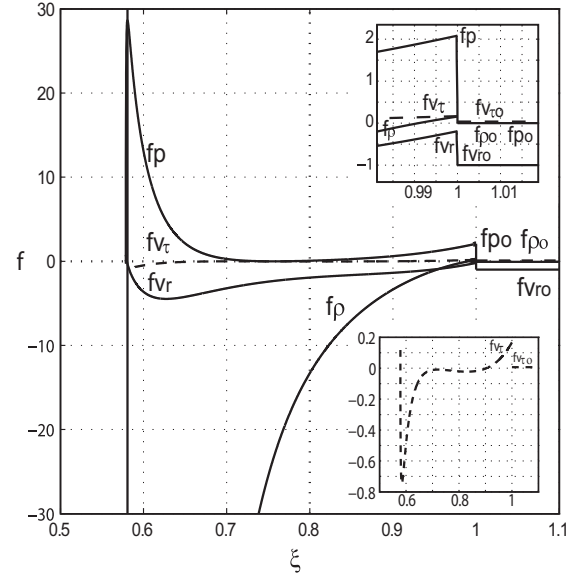}
\caption{The real parts of the eigenfunctions for the density, radial velocity and pressure perturbations (solid lines), the tangential velocity (dashed line) and the external vorticity perturbations (lines for $\xi>1$) in the case of the shell-like flow ($l=10$, $s=0.1304 i$, $\omega=2.7$, $\gamma=4/3$). The imaginary parts are identically zero.}\label{eigenfunctions}
\end{figure}

Given the amplitudes and frequency of the external perturbations, the structure of the perturbed flow inside the shock wave is determined by solving
in the interval $\xi_{in}\le\xi\le 1$, $\xi_{in}\le 0$ the system of linear ODEs \eqref{eqv1}-\eqref{eqv3},\eqref{eqv5} for unknown variables $f_{v_{\rho}}(\xi)$, $f_{v_{\xi}}(\xi)$, $f_{v_{\tau}}(\xi)$, $f_ {p}(\xi)$ with the outer \eqref{Renken6}-\eqref{Renken9} and inner \eqref{obolochkainer}, \eqref{In1bez} or \eqref{In2bez} boundary conditions. These unknown variables represent the eigenfunctions for the problem while $\eta$ is now an eigenvalue. We find the root $\eta$ by multiple shooting method terminating iterations upon reaching relative precision at least $10^{-4}$.
In the case of the filled cavity the problem may become stiff; this dictates sometimes a need for much better accuracy to satisfy the inner boundary condition.
Figure ~\ref{eigenfunctions} demonstrates the typical behavior of the eigenfunctions in the case of shell-like flow.

To estimate the possible amplification of perturbations by the shock wave we introduce the following coefficients of

\noindent density amplification
\begin{equation}\label{R}
R=\frac{|f_{\rho_s}|}{|f_{\rho_0}|}\,,
\end{equation}
velocity amplification
\begin{equation}\label{Xi}
\Xi=\frac{|\mathbf{f}_{v_s}|}{|\mathbf{f}_{v_0}|}\,,
\end{equation}
and vorticity amplification
\begin{equation}\label{Omega}
\Omega=\frac{|{\rm curl}\,\mathbf{f}_{v_s}|}{|{\rm curl}\,\mathbf{f}_{v_0}|}\,.
\end{equation}
Accounting for an explicit form for the curl of the vector $\delta \mathbf{v}$ with zero binormal component in natural coordinates
\begin{equation}\label{curl_deltav}
\nabla \times \delta \mathbf{v} =
\mathbf{e}_b \left(  \frac{ d \delta v_{\tau}}{d\xi} + \frac{\delta v_{\tau}}{\xi}  - \frac{\delta v_{\xi}}{\xi} \right),
\end{equation}
we can write out the vorticity amplification coefficient in full as
\begin{equation}\label{Omega_explicit}
\Omega=
\frac{ | \xi \frac{d f_{v_{\tau }}}{d \xi}\big|_s + f_{v_{\tau s}} - f_{v_{\xi s}} | }{ |f_{v_{\xi 0}} - f_{v_{\tau 0}}| }.
\end{equation}
All these coefficients of amplifications are ratios of amplitudes and time-constant by definition.

\begin{figure}[t]
\epsscale{1.16}
\plotone{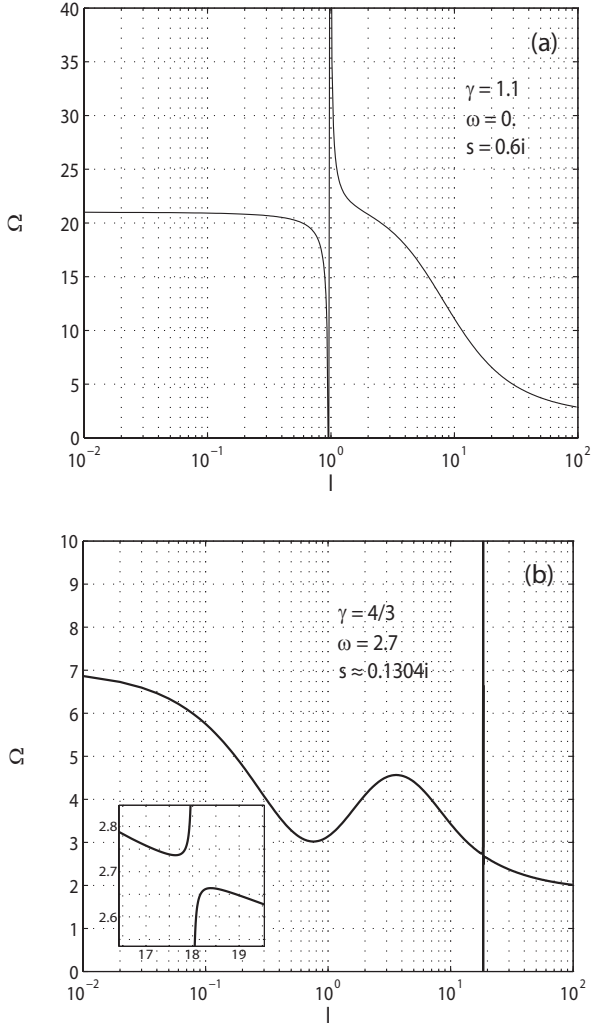}
\caption{The vorticity amplification coefficient (Eq.~\eqref{Omega}) as a function of the orbital wavenumber $l$ for the uniform (upper plot) and the nonuniform background (lower plot).}\label{vorticity_ampl}
\end{figure}

\begin{figure}[t]
\epsscale{1.16}
\plotone{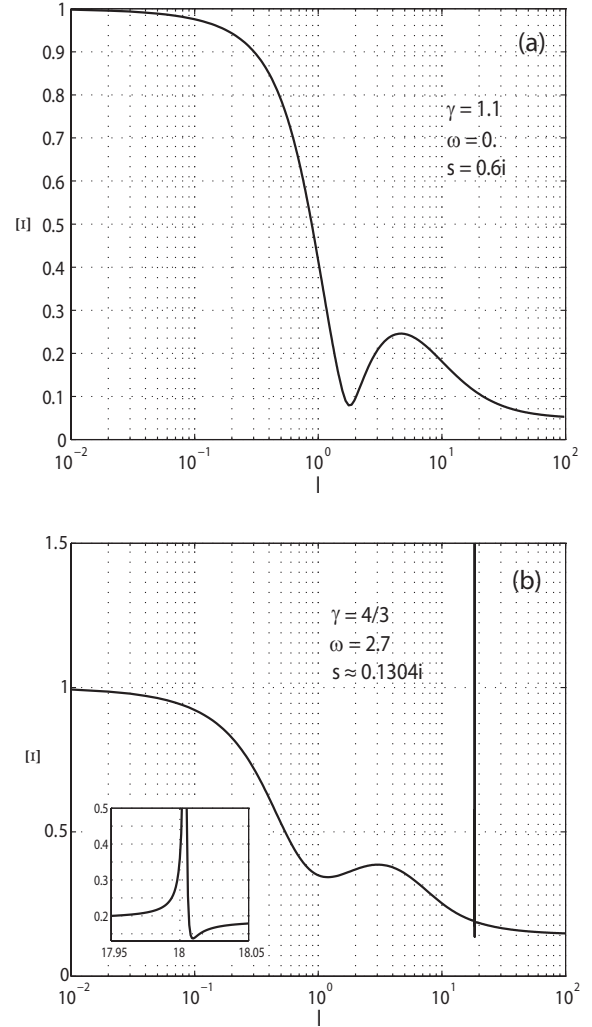}
\caption{The same as in Fig.~\ref{vorticity_ampl} but for the velocity amplification (Eq.~\eqref{Xi}).}\label{velocity_ampl}
\end{figure}

First let us consider the vortex mode whose frequency is given by Equation ~\eqref{statrel3}.

The simplest perturbation of the velocity field is a uniform motion $\delta \mathbf{v}_0={\rm const}$ which contains only a dipole component in spherical harmonics expansion and corresponds to a steady drift of the whole shock wave. Such a flow is purely irrotational, the denominator in \eqref{Omega} vanishes but the numerator remains finite. We can observe this effect as a
resonance at $l=1$ in the case of the uniform background (Figure ~\ref{vorticity_ampl}). The same effect exists of course in the case of any background which involves $\delta\rho_0\ne 0$ however, but since we consider the vortex and the entropy modes separately, we do not observe this resonance in other figures.

Let us remember that the curl of velocity for any harmonic \eqref{e1}
is expressed as a product of an amplitude depending on $\xi$ and the angular part  $l Y_{lm}$.
Thus the coefficient \eqref{Omega} actually presents the amplitudes ratio irrespective of $l Y_{lm}$. The limit $l\to 0$ thus provides an evaluation of indeterminate form as zero divide zero because the vorticities themselves $\delta \mathbf{v}_0(l=0)$ and $\delta \mathbf{v}_s(l=0)$ vanish.

From Figures ~\ref{vorticity_ampl},\ref{velocity_ampl} we see that vorticity intensifies substantially up to 20 times within the wavenumbers range under consideration out of resonance while the velocity fluctuations are damped.
The reason for vorticity amplification is that the rippled shock front generates the shear flow.
While the radial velocity perturbations are strongly suppressed the tangential perturbations are excited.
For instance, for the shortwave perturbation $l=10$ whose eigenfunctions are depicted in Figure~\ref{eigenfunctions}, the preshock velocity perturbations are radially dominated due to relation \eqref{svas} ($f_{v_{\xi 0}}=1.0, f_{v_{\tau 0}}=0.006$) but the amplitudes of the postshock velocity components equalize ($f_{v_{\xi s}}=0.19, f_{v_{\tau s}}=0.17$).
Along with that the tangential perturbation derivative $df_{v_{\tau}}/d\xi$ undergoes a steep rise from zero before the shock to 3.1 behind the shock (see right bottom inset in Figure~\ref{eigenfunctions}) building up vorticity significantly (see Equation~\eqref{Omega_explicit}).

The most intriguing effect is the resonance that we find in the case $\omega=2.7$ at the resonant wavenumber $l=l^*$ given by Equation \eqref{Reson_point_vortex_om27}. The insets in Figures ~\ref{vorticity_ampl},\ref{velocity_ampl} show that the resonance width is rather small and equals $\Delta l\approx 0.4$ for vorticity and $< 0.1$ for velocity amplification. Here the resonance width is defined as a range of wavenumbers $l$ in which the frequency differs from the `mean' values by the value of $\Delta {\rm Im}\,s \sim 0.1$. Strictly speaking, according to the linear theory of oscillations the width of resonance in the dissipationless case is negligible so long as the width is usually defined as the interval at which the frequency differs no more than twice from the maximum value.

\begin{figure}
\epsscale{1.16}\plotone{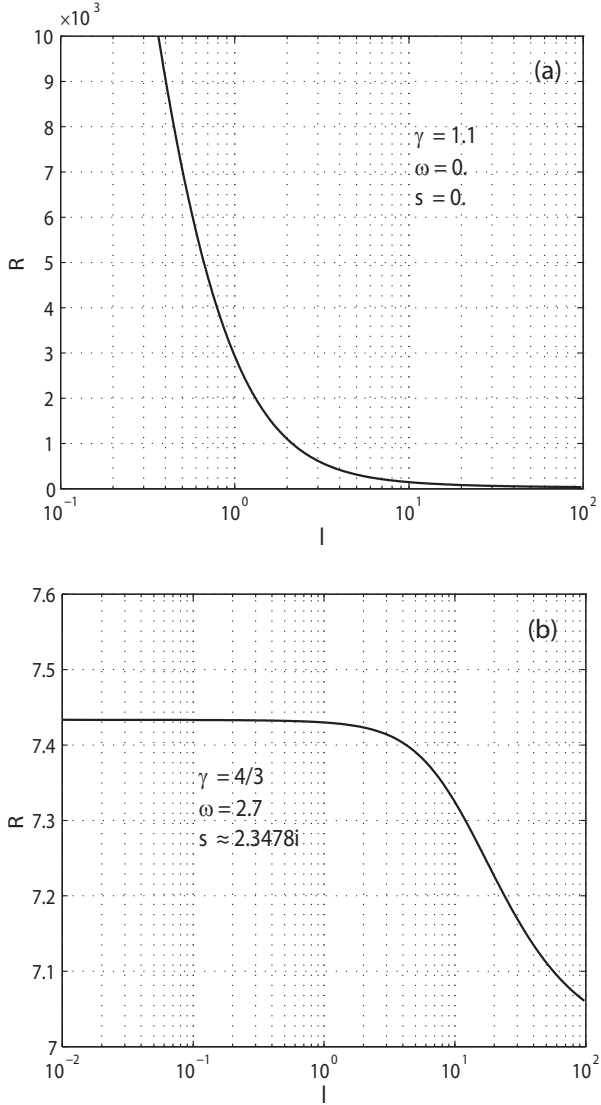}\caption{The same as in Fig.~\ref{vorticity_ampl} but for the density amplification (Eq.~\eqref{R}).}\label{density_ampl}
\end{figure}

The density perturbations in the entropy mode are amplified out of resonance only in the case of the accelerating shell blast wave ($\omega=2.7$) whereas they are damped by decelerating solid blast wave ($\omega=0$) (Figure ~\ref{density_ampl}).

Resonance prescribed by Equation ~\eqref{Reson_point_entropy_om0} at $l=0,\ \omega=0$ has a greater width: $\Delta l=2$. This resonance describes self-excitation of radial fluctuations of the shock at negligible external density perturbations and takes place for arbitrary $\omega$ because dispersion dependencies always comprise the point $(s=0,\ l=0)$. Suppose the energy of explosion undergoes small perturbation $E_0\to E_0+\delta E_0$. Then the relative displacement of the shock front changes over time as
\begin{equation}\label{deltaRs_Rs}
\begin{split}
\eta &= \frac{R_s(E_0+\delta E_0,t)-R_s(E_0,t)}{R_s(E_0,t)} \\
     &= \frac{(E_0+\delta E_0)^{\frac{1}{5-\omega}}t^{\frac{2}{5-\omega}}-E_0^{\frac{1}{5-\omega}} t^{\frac{2}{5-\omega}}}{E_0^{\frac{1}{5-\omega}} t^{\frac{2}{5-\omega}}}\propto t^0,
\end{split}
\end{equation}
that is, according to Equation \eqref{eta} the spherical perturbation will have zero frequency.

\section{DISCUSSION AND CONCLUSION}

Our linear analysis validates an idea that the blast wave can amplify both density and vorticity perturbations in the turbulized interstellar medium through
resonant shock-turbulence interaction. This result is consistent with analytic and numerical studies \citep{Lee97,Jamme2002,Wouchuk} predicting vorticity amplification by a distorted shock front. As distinct from the aforementioned works we study the non-local mechanism of amplification when the wave lengths $\lambda$ of perturbations are not negligible compared with the shock wave size $R_s$. This singles out the resonant nondimensional wavenumbers at macro- ($l= {2\pi R_s}/{\lambda} \sim 1$), meso- ($l \approx 18$) and microscopic ($l > 200$) scales. Resonant mesoscales are surprisingly very close to the observed scales ($l \approx 15$) of corrugation structures in the remnant SNR 0509-67.5.

On the other hand, there are some factors negating the effectiveness of the resonant mechanism.
First, the resonance width is small (typically $\Delta l \ll 1$) and the resonance point can be clamped between two nearly integer points $l$. Since the observed orbital wavenumbers $l$ must be integer values, the amplification for these will not be too large.

The resonance range can grow in size however with dissipation increase. The linear theory of oscillations predicts that the resonance width of frequencies is proportional to the doubled decay coefficient. Taking into account that in our self-similar problem the frequency is defined as $s\log(t/t_0) /t$  and the decay has to be stipulated by the turbulent viscosity we find that  $\Delta {\rm Im}\,s \log(t/t_0)/t \approx 2 \nu_{turb}/ \lambda^2$, where $\nu_{turb}$ is the turbulent kinematic viscosity.
The length of the perturbation inside the shock wave can be estimated as $\lambda\approx 2\pi R_s/\sqrt{l^2+n^2}$ while the kinematic viscosity approximately is $\nu_{turb}\approx \frac{1}{3}l_{turb} c_s$, where $c_s$ is the postshock sound velocity and $l_{turb}$ is the characteristic scale of a turbulent vortex. Expressing the resonant width in terms of the wavenumbers for the resonant case considered in our model, $n=0$ and $l=l^*$, we get
\begin{equation}\label{Res_width_1}
  \frac{\Delta l}{l^*} \approx \frac{l^* l_{turb} c_s t}{6\pi^2 R_s^2 \log(t/t_0)} \left( \frac{d {\rm Im}\,s}{d l}\bigg|_{l^*} \right)^{-1}.
\end{equation}
The postshock sound velocity is determined through the relations \eqref{R_s},\eqref{vprho_rs}:
\begin{equation}\label{c_s_estim}
  c_s = Z \frac{R_s}{t},
\end{equation}
where
\begin{equation}\label{Z}
  Z = \frac{2\sqrt{2\gamma(\gamma-1)}}{(5-\omega)(\gamma+1)}.
\end{equation}
For $\gamma=1.1$ and $\omega=0$ we have $Z\approx 0.089$, for $\gamma=4/3$, $\omega=2.7$ the coefficient $Z$ is something like $0.35$.
Since the propagating shock embraces growing scales we estimate  the turbulent vortex scale $l_{turb}$ as proportional to the shock front radius $R_s$ with a fixed constant of proportionality $\xi_{turb}$.
Collecting all estimates we finally obtain
\begin{equation}\label{Res_width_2}
  \frac{\Delta l}{l^*} \approx \frac{l^* \xi_{turb} Z}{6\pi^2  \log(t/t_0)} \left( \frac{d {\rm Im}\,s}{d l}\bigg|_{l^*} \right)^{-1}.
\end{equation}
The numerical calculations (Fig.~\ref{spectrg43om27}) show that the derivative of the dispersion relation $\frac{d {\rm Im}\,s}{d l}\big|_{l^*}$ amounts to $\sim 0.1$ at the resonant point $l^*=18$. Assuming $\xi_{turb}\approx 0.1$ we find
\begin{equation}\label{Res_width_3}
  \frac{\Delta l}{l^*} \log(t/t_0) \approx  0.025,
\end{equation}
which is close to the width of resonance found numerically (Figs.~\ref{vorticity_ampl}b,\ref{velocity_ampl}b).

The estimates adduced show that the resonance range subject to the existence of dissipation is small but widens and can be not vanishingly small if the resonance occurs at a more gently sloping dispersion curve ($\frac{d {\rm Im}\,s}{d l}\big|_{l^*} \to 0$) and for higher order harmonics $l^*$ and $n$.

Secondly, to excite resonant oscillations noticeably the amplitude of resonant harmonics in the spectrum of external perturbations should be non-negligible.
Suppose turbulent fluctuations energy distribution over discrete orbital quantum numbers $l$ has a power dependence like Kolmogorov law $f_l^2\sim l^{-5/3}$, and if we are fortunate it starts from $l_{min}$ where $l_{min}$ is close to the resonant number $l^*=18$, then the relative  square of amplitude of the $l=l_{min}=18$ mode amounts to $f^2_{l_{min}} / ( \sum_{l=l_{min}}^{\infty}{f_l^2}) \approx 3.7\%$ of the total energy of fluctuations.
The estimate can be reduced significantly if $l_{min}<l^*$.

To clarify the real effectiveness of the resonant mechanism of amplification a direct numerical simulation is required.

\acknowledgments
The work was supported by the RFBR grant 15-42-02682-r-povolzhie\_a and 
 by the State agreement 14.B37.21.0915 of the Russian Ministry for Education and Research.
Andrey~Zankovich expresses his gratitude to the Dynasty Foundation of Dmitry Zimin for the scholarship.

\end{document}